# Strong interlayer coupling in van der Waals heterostructures built from single-layer chalcogenides

Hui Fang[a,b], Corsin Battaglia[a,b], Carlo Carraro[c], Slavomir Nemsak[b,d], Burak Ozdol[e,f],

Jeong Seuk Kang[a,b], Hans A. Bechtel[g], Sujay B. Desai[a,b], Florian Kronast[h], Ahmet A. Unal[h],

Giuseppina Conti[b,d], Catherine Conlon[b,d], Gunnar K. Palsson[b,d], Michael C. Martin[g],

Andrew M. Minor[e,f], Charles S. Fadley[b,d], Eli Yablonovitch[a,b,1], Roya Maboudian[c], and Ali Javey[a,b,1]

[a]Electrical Engineering and Computer Sciences, University of California, Berkeley, CA 94720, USA.

[b]Materials Sciences Division, Lawrence Berkeley National Laboratory, Berkeley, CA 94720, USA.

[c]Chemical and Biomolecular Engineering, University of California, Berkeley, CA 94720, USA.

[d]Physics, University of California, Davis, CA 95616, USA.

[e]Materials Science and Engineering, University of California, Berkeley, CA 94720, USA.

[f]National Center for Electron Microscopy, Lawrence Berkeley National Laboratory, Berkeley, CA 94720

[g]Advanced Light Source, Lawrence Berkeley National Laboratory, Berkeley, CA 94720, USA.

[h]Helmholtz-Zentrum Berlin für Materialien und Energie GmbH, Albert-Einstein-Straße 15, D-12489 Berlin, Germany.

[1]Correspondence should be addressed to eliy@eecs.berkeley.edu or ajavey@eecs.berkeley.edu.

**Semiconductor heterostructures are the fundamental platform for many important device applications such as lasers, light-emitting diodes, solar cells and high-electron-mobility transistors. Analogous to traditional heterostructures, layered transition metal dichalcogenide (TMDC) heterostructures can be designed and built by assembling individual single-layers into functional multilayer structures, but in principle with atomically sharp interfaces, no interdiffusion of atoms, digitally controlled layered components and no lattice parameter constraints. Nonetheless, the optoelectronic behavior of this new type of van der Waals (vdW) semiconductor**

**heterostructure is unknown at the single-layer limit. Specifically, it is experimentally unknown whether the optical transitions will be spatially direct or indirect in such hetero-bilayers. Here, we investigate artificial semiconductor heterostructures built from single-layers WSe$_2$ and MoS$_2$. We observe a large Stokes-like shift of ~100meV between the photoluminescence peak and the lowest absorption peak that is consistent with a type II band alignment having spatially *direct* absorption but spatially *indirect* emission. Notably, the photoluminescence intensity of this spatially indirect transition is strong, suggesting strong interlayer coupling of charge carriers. This coupling at the hetero-interface can be readily tuned by inserting dielectric layers into the vdW gap, consisting of hexagonal BN (h-BN). Consequently the generic nature of this interlayer coupling provides a new degree of freedom in band engineering and is expected to yield a new family of semiconductor heterostructures having tunable optoelectronic properties with customized composite layers**.

Two-dimensional (2D) layered TMDC semiconductors such as MoS$_2$ and WSe$_2$ have established themselves as strong contenders for next generation electronics and optoelectronics (1-6) and are promising building blocks for novel semiconductor heterostructures (7-11). Conventional heterostructures are mainly based on group IV, III-V or II-VI semiconductors with covalent bonding between atoms at the hetero-interface. Due to atomic interdiffusion during growth, the resulting atomic-scale interface roughness and composition variation at the hetero-interface inevitably smear the density of states profile and consequently compromise the performance of these heterostructures especially as the film thicknesses are reduced towards a single atomic layer. In addition, the choice of material components for conventional heterostructures is strongly dictated by lattice mismatch.

In TMDCs on the other hand, individual layers are held together by vdW forces, without surface dangling bonds (12). Semiconductor heterostructures built up from mono-layer TMDCs would in principle offer atomically regulated interfaces and thereby sharp band edges. Theoretical studies have predicted different electronic structures and optical properties from TMDC hetero-bilayers (13-17), however, to date there has been no experimental results. While previous experimental efforts have focused on graphene based layered heterostructures (8-11, 18-26), we present here an experimental study on the electronic interlayer interaction in a heterostructure built from two single-layer TMDC semiconductors, namely $MoS_2$ and $WSe_2$. The hetero-bilayers are characterized by transmission electron microscopy, x-ray photoelectron microscopy, electron transport studies and optical spectroscopy, to elucidate the band alignments, optoelectronic properties and the degree of the electronic layer coupling in this novel material system.

The fabrication of $WSe_2/MoS_2$ hetero-bilayers was realized by stacking individual mono-layers on top of each other (see Methods in Supporting Information for details). Figure 1A shows an illustration of the hetero-bilayer, and Fig. 1B displays the corresponding optical microscope image of a $WSe_2/MoS_2$ hetero-bilayer on a Si substrate with 260nm thermally grown $SiO_2$. Owing to the 3.8% lattice mismatch, estimated from the bulk lattice constants (12), as well as the unregulated, but in principle controllable angular alignment ($\phi$) between the constituent layers, the heterostructure lattice forms a moiré pattern, clearly visible in the high resolution transmission electron microscopy (HRTEM) image in Fig. 1C. The HRTEM image displays the boundary region between single-layer MoS$_2$ and the WSe$_2$/MoS$_2$ hetero-bilayer. While MoS$_2$ exhibits a simple hexagonal lattice, the heterostructure shows moiré fringes with a spatial envelope periodicity on the order of 4-6× the lattice constants of WSe$_2$ (or MoS$_2$). Inspection of the diffraction pattern in Fig. 1D along the [001] zone axis reveals that in this particular sample the two hexagonal reciprocal lattices are rotated by $\phi$=12.5° with respect to each other and there is negligible strain in the two constituent layers (Supporting Information). The alignment of the two lattices can also be examined with a fast Fourier transform of the two zoomed-in TEM images in Fig. 1C (Supporting Fig. S1). The absence of strain in the constituent layers of the WSe$_2$/MoS$_2$ hetero-bilayer is also confirmed by Raman spectroscopy (Supporting Fig. S2), which show that the in plane vibration modes of both WSe$_2$ and MoS$_2$ maintain their corresponding positions before and after transfer.

To shed light on the electronic structure of the WSe$_2$/MoS$_2$ heterostack, we performed x-ray photoelectron spectroscopy (XPS). Specifically, we used a photoemission electron microscope (PEEM) with a high spatial resolution of 30nm to discriminate between photoelectrons emitted from the WSe$_2$ single-layer, MoS$_2$ single-layer and the WSe$_2$/MoS$_2$ hetero-bilayer, as illustrated in Fig. 2A (see Supporting Fig. S3 for details). In addition, by looking at the core-level photoelectrons, we achieved elemental and electronic selectivity that allows us to probe photoelectrons originating from the top layer of the hetero-bilayer and to directly quantify the potential difference between the WSe$_2$ layer in the hetero-stack with respect to the WSe$_2$ single-layer reference on the substrate. As shown in Fig. 2B, a peak shift of about -220 meV in binding energy (or +220 meV in kinetic energy) is evident in the W 4f core levels of the hetero-bilayer as compared to WSe$_2$ single layer. The direction of the peak shift is consistent with a negative net charge on the WSe$_2$ in the WSe$_2$/MoS$_2$ hetero-bilayer. On the other hand, a shift of +190 meV is observed in the Mo 3d core levels of the WSe$_2$/MoS$_2$ in Fig. 2C. Our PEEM results therefore indicate that the WSe$_2$ layer has a negative net charge, while the MoS$_2$ layer has a positive net charge as a result of Contact Potential. The hetero-bilayer can essentially be interpreted as being a two-dimensional dipole, an atomically thin parallel plate capacitor with vdW gap with a built-in potential up to 400 mV, originating from the work function difference induced charge transfer between the two constituent single layers. The latter interpretation is also consistent with the p- and n-type character of WSe$_2$ and MoS$_2$, respectively (ref. (2, 3)).

To investigate the optoelectronic properties of the WSe$_2$/MoS$_2$ hetero-bilayer, we used photoluminescence (PL) and absorption spectroscopy. It is known that both single-layer WSe$_2$ and MoS$_2$ exhibit direct band gaps, while their bulk and homo-bilayer counterparts are indirect (1, 27). In agreement with previous work we observe strong excitonic PL peaks at 1.64 eV and 1.87 eV for single-layer WSe$_2$ and MoS$_2$ respectively (Fig. 3A). Note that single-layer WSe$_2$ shows a 10-20× higher PL intensity than single-layer MoS$_2$, a result consistent with ref. (28). For the WSe$_2$/MoS$_2$ hetero-bilayer, we observe a peak at 1.55 eV, lying interestingly at a lower energy than for the two constituent single layers, as shown in Fig. 3A, (with intensity ~1.5× higher than for single-layer MoS$_2$). The appearance of a peak at such low energy was observed consistently for multiple (>10) samples, with peak energies ranging from 1.50 to 1.56 eV (Supporting Fig. S4). This distribution is attributed to sample-to-sample variations in interface quality and/or alignment angle $\phi$. Of value in optoelectronics, an Urbach tail inverse slope, corresponding to the band-edge sharpness of ~30 meV/dec is extracted from the PL spectra (29, 30) (Supporting Fig. S5). The steep tail slopes of our hetero-bilayer proves that high-quality heterostructures with sharp band edges can be built at the single-layer limit using TMDC building blocks which is a unique feature of this material system.

The nature of the photoluminescence of the $WSe_2/MoS_2$ hetero-bilayer is intriguing. To better understand the electronic structure of the hetero-bilayer, we performed absorption measurements in the near-infrared and visible part of the spectrum using synchrotron light shown as dashed lines in Fig. 3B. The $WSe_2/MoS_2$ hetero-bilayer shows a first absorption peak at 1.65 eV and a second peak at 1.91 eV. These peaks closely coincide with the absorption peaks of single-layer $WSe_2$ and $MoS_2$, respectively. Interestingly, comparing the absorption spectra with the normalized PL data shown in Fig. 3B, we note that the hetero-bilayer exhibits a striking ~100 meV shift between the PL and absorbance peaks. This large Stokes-like shift is consistent with a spatially indirect transition in a staggered gap (type II) heterostructure (31) (as shown in Fig. 3C). Our hetero-bilayers share certain similarities with organic semiconductor heterostructures in which donor and acceptor layers are also bound by weak intermolecular vdW forces (32). Similar to the optical processes in organic heterostructures, photons are absorbed in single layer $WSe_2$ and single layer $MoS_2$, generating excitons in both layers. Photo-excited excitons then relax at the $MoS_2/WSe_2$ interface, driven by the band-offset as shown in Fig. 3C. That band offset is also consistent with the measured built-in electric field from PEEM. Owing to the energy lost to the band offset (Fig. 3C), the PL excitonic peak energy is lower than the excitonic band gaps of either material component.

This 100meV shift, may be a balance between conduction band offset between the two monolayers versus diminished exciton binding energy associated with being spatially indirect. Note that in the hetero-bilayer, we observe only a weak luminescence signal at the energies corresponding to the excitonic band gaps of single layer $MoS_2$ and $WSe_2$, suggesting that the large majority of the photo-excited carriers are relaxed at the interface producing the highest luminescence for the spatially indirect recombination process.

To fine tune the interlayer interaction in the $WSe_2/MoS_2$ hetero-bilayer, single- and few-layer sheets of h-BN spacer layers were inserted into the vdW gap (Fig. 3D) using the same transfer technique. Fig. 3E shows the normalized PL of hetero-stacks with single- and tri-layer h-BN spacers. Interlayer spatially indirect recombination becomes negligible for the sample with a tri-layer h-BN spacer, as indicated by both the position and the intensity of the peak at 1.64 eV (Fig. 3E and Supporting Fig. S6), which are nearly the same as for single-layer $WSe_2$. On the other hand, a single layer of h-BN does not fully suppress the interlayer interaction between $WSe_2$ and $MoS_2$. The results demonstrate that the interlayer coupling can be readily tuned by intercalation of dielectric layers, and provide yet another degree of control in the vdW heterostructure properties.

Finally, we explored the carrier transport along the hetero-bilayer interface. A single flake consisting of single-layer $WSe_2$ and $MoS_2$, and an overlapping hetero-bilayer was made via the transfer process. The flake was dry etched into a long ribbon (Fig. 4A). A corresponding PL peak energy map is shown at the right edge of Fig. 4B, further depicting the ribbon structure by color coding of the luminescence energy. Multiple source/drain (S/D) metal electrodes were then fabricated by electron beam lithography and lift-off on each region of the ribbon (see Figs. S7 and S8 for details). The Si/SiO$_2$ substrate serves as the global back gate, with 260 nm gate oxide thickness). As expected single-layer MoS$_2$ and WSe$_2$ devices exhibit *n* and *p*-channel characteristics, respectively, (Supporting Fig. S9), consistent with previous reports (2, 3). On the other hand, the device consisting of one contact on the mono-layer WSe$_2$ and the other on mono-layer MoS$_2$, with the two layers overlapping in the central region (Fig. 4B) exhibits a distinct rectifying behavior (Fig. 4C and Fig. S10), consistent with type II band-alignment of the heterobilayer. The rectification provides additional evidence for electrical coupling and proper Contact Potential between the two constituent layers. This behavior is consistent with previous work on TMDC/nanotubes (33) and TMDC/III-V heterostructures (34), which had shown that electrically active vdW interfaces can be achieved from TMDC components. The work here highlights the ability to engineer a novel class of electronic and optoelectronic devices by vdW stacking of the desired layered chalcogenide components with molecular-scale thickness control.

In summary, we have fabricated and characterized an artificial vdW heterostructure by stacking mono-layer transition metal dichalcogenide building blocks and achieved electronic coupling between the two 2D semiconductor constituents. Strong PL with a large Stokes-like shift was observed from the WSe$_2$/MoS$_2$ hetero-bilayer, consistent with spatially indirect luminescence from a type II heterostructure. We anticipate that our result will trigger subsequent studies focused on the bottom-up creation of new heterostructures by varying chemical composition, interlayer spacing and angular alignment. In addition the focus will be on the fabrication of vdW semiconductor heterostructure devices, with tuned optoelectronic properties from customized single layer components. Particularly, electroluminescene efficiency of vdW heterostructures needs to be explored experimentally to examine their viability for use as nanoscale light emitting/lasing devices.

**Acknowledgements:**

Materials characterization of this work was supported by the Director, Office of Science, Office of Basic Energy Sciences, Materials Sciences and Engineering Division of the U.S. Department of Energy under Contract No. DE-AC02-05CH11231. The device fabrication and characterization was supported by NSF $E^3S$ center. The TEM work was performed at the National Center for Electron Microscopy, Lawrence Berkeley National Laboratory, which is supported by the U.S. Department of Energy under Contract # DE-AC02-05CH11231. The absorption measurements were performed at the Advanced Light Source, Lawrence Berkeley National Laboratory. C. Carraro. and R. M. acknowledge support from the National Science Foundation under grant # EEC-0832819 (Center of Integrated Nanomechanical Systems).

**Author Contributions**

H.F., C. Carraro., R.M. and A.J. conceived the experiments. H.F., C. Carraro., S.N., B.O., J.S.K., H.A.B., S.B.D., F.K., A.A.U., C. Conlon., G.K.P. and M.C.M. carried out the experiments. H.F., C.B., C. Carraro, S.N., B.O., G.C., A.M., C.S.F., E.Y., R.M. and A.J. contributed to the data analysis. H.F., C.B. and A.J. wrote the paper while all authors provided feedback.

**References**

1. Mak KF, Lee C, Hone J, Shan J, & Heinz TF (2010) Atomically Thin $MoS_2$: A New Direct-Gap Semiconductor. *Phys Rev Lett* 105(13):136805.

2. Radisavljevic B, Radenovic A, Brivio J, Giacometti V, & Kis A (2011) Single-layer $MoS_2$ transistors. *Nat Nanotechnol* 6(3):147-150.

3. Fang H, *et al.* (2012) High-Performance Single Layered $WSe_2$ p-FETs with Chemically Doped Contacts. *Nano Lett* 12(7):3788-3792.

4. Wang H, *et al.* (2012) Integrated circuits based on bilayer $MoS_2$ transistors. *Nano Lett* 12(9):4674-4680.

5. Zeng H, Dai J, Yao W, Xiao D, & Cui X (2012) Valley polarization in $MoS_2$ monolayers by optical pumping. *Nat Nanotechnol* 7(8):490-493.

6. Jones AM, *et al.* (2013) Optical Generation of Excitonic Valley Coherence in Monolayer $WSe_2$. *Nature Nanotechnol* 8:634-638.

7. Geim A & Grigorieva I (2013) Van der Waals heterostructures. *Nature* 499(7459):419-425.

8. Yu WJ, *et al.* (2012) Vertically stacked multi-heterostructures of layered materials for logic transistors and complementary inverters. *Nat Mater* 12(3):246-252.

9. Georgiou T, *et al.* (2012) Vertical field-effect transistor based on graphene-$WS_2$ heterostructures for flexible and transparent electronics. *Nat Nanotechnol* 8(2):100-103.

10. Yu WJ, *et al.* (2013) Highly efficient gate-tunable photocurrent generation in vertical heterostructures of layered materials. *Nat Nanotechnol* 8(12):952-958.

11. Britnell L, *et al.* (2013) Strong light-matter interactions in heterostructures of atomically thin films. *Science* 340(6138):1311-1314.

12. Wilson J & Yoffe A (1969) The transition metal dichalcogenides discussion and interpretation of the observed optical, electrical and structural properties. *Adv Phys* 18(73):193-335.

13. Terrones H, López-Urías F, & Terrones M (2013) Novel hetero-layered materials with tunable direct band gaps by sandwiching different metal disulfides and diselenides. *Sci Rep* 3:1549.

14. Kang J, Li J, Li S-S, Xia J-B, & Wang L-W (2013) Electronic structural Moiré pattern effects on $MoS_2$/$MoSe_2$ 2D heterostructures. *Nano Lett.* 13(11):5485-5490.

15. Kośmider K & Fernández-Rossier J (2013) Electronic properties of the $MoS_2$-$WS_2$ heterojunction. *Phys Rev B* 87(7):075451.

16. Komsa H-P & Krasheninnikov AV (2013) Electronic structures and optical properties of realistic transition metal dichalcogenide heterostructures from first principles. *Phys Rev B* 88(8):085318.

17. Gong C, *et al.* (2013) Band alignment of two-dimensional transition metal dichalcogenides: Application in tunnel field effect transistors. *Appl Phys Lett* 103(5):053513.

18. Dean C, *et al.* (2010) Boron nitride substrates for high-quality graphene electronics. *Nat Nanotechnol* 5(10):722-726.

19. Ponomarenko L, *et al.* (2011) Tunable metal-insulator transition in double-layer graphene heterostructures. *Nat Phys* 7(12):958-961.

20. Haigh S, *et al.* (2012) Cross-sectional imaging of individual layers and buried interfaces of graphene-based heterostructures and superlattices. *Nat Mater* 11:764-767.

21. Britnell L, *et al.* (2012) Field-effect tunneling transistor based on vertical graphene heterostructures. *Science* 335(6071):947-950.

22. Gorbachev R, *et al.* (2012) Strong Coulomb drag and broken symmetry in double-layer graphene. *Nat Phys* 8:896-901.

23. Hunt B, *et al.* (2013) Massive Dirac fermions and Hofstadter butterfly in a van der Waals heterostructure. *Science* 340(6139):1427-1430.

24. Ponomarenko L, *et al.* (2013) Cloning of Dirac fermions in graphene superlattices. *Nature* 497(7451):594-597.

25. Dean C, *et al.* (2013) Hofstadter's butterfly and the fractal quantum Hall effect in moire superlattices. *Nature* 497(7451):598-602.

26. Yankowitz M, *et al.* (2012) Emergence of superlattice Dirac points in graphene on hexagonal boron nitride. *Nat Phys* 8(5):382-386.

27. Zeng H, *et al.* (2013) Optical signature of symmetry variations and spin-valley coupling in atomically thin tungsten dichalcogenides. *Sci Rep* 3:1608.

28. Zhao W, *et al.* (2012) Evolution of electronic structure in atomically thin sheets of $WS_2$ and $WSe_2$. *ACS nano* 7(1):791-797.

29. Van Roosbroeck W & Shockley W (1954) Photon-radiative recombination of electrons and holes in germanium. *Phys Rev* 94(6):1558.

30. Kost A, Lee H, Zou Y, Dapkus P, & Garmire E (1989) Band-edge absorption coefficients from photoluminescence in semiconductor multiple quantum wells. *Appl Phys Lett* 54(14):1356-1358.

31. Wilson BA (1988) Carrier dynamics and recombination mechanisms in staggered-alignment heterostructures. *IEEE J Quant Electron* 24(8):1763-1777.

32. Li G, Zhu R, & Yang Y (2012) Polymer solar cells. *Nat Photon* 6(3):153-161.

33. Jariwala D, *et al.* (2013) Gate-tunable carbon nanotube–MoS$_2$ heterojunction p-n diode. *Proc Nat Acad Sci* 110(45):18076-18080.

34. Chuang S, *et al.* (2013) Near-ideal electrical properties of InAs/WSe$_2$ van der Waals heterojunction diodes. *Appl Phys Lett* 102(24):242101.

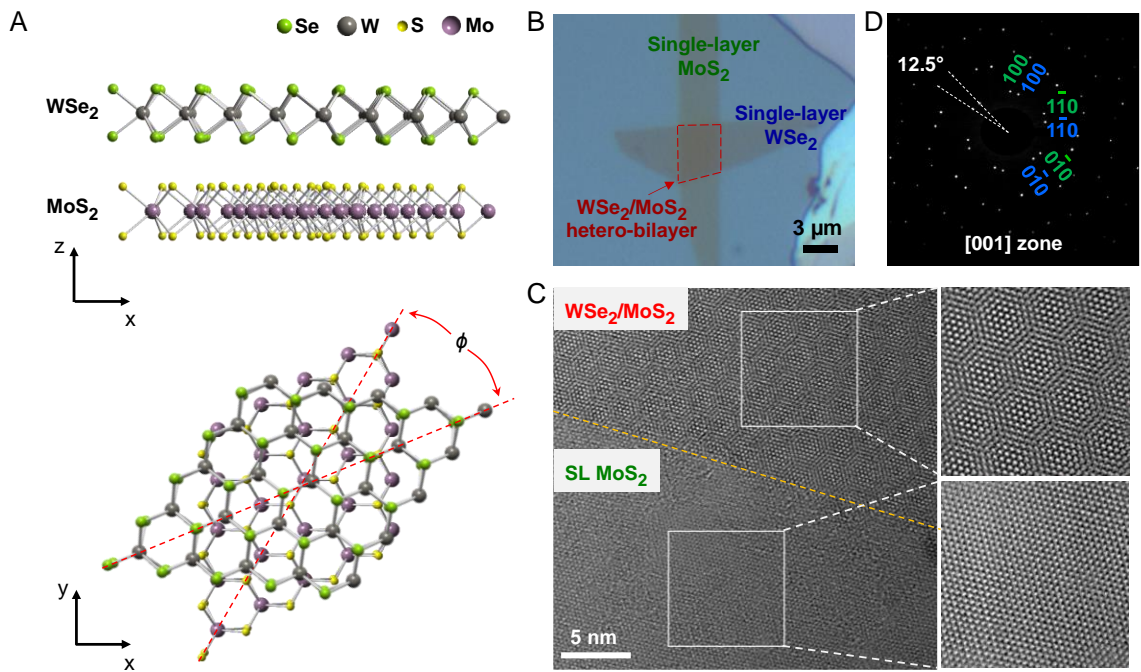

**Figure 1.** WSe$_2$/MoS$_2$ hetero-bilayer illustration, optical image, and TEM images. (A) Atomistic illustrations of the heterostructure of single-layer (SL) WSe$_2$ on single-layer MoS$_2$ with their respective lattice constants and a misalignment angle $\phi$. (B) Optical microscope image of a WSe$_2$/MoS$_2$ hetero-bilayer on a Si/SiO$_2$ substrate (260nm SiO$_2$). (C) High resolution TEM images of a boundary region of single-layer MoS$_2$ and the hetero-bilayer, showing the resulting Moiré pattern. (D) The electron diffraction pattern of the hetero-bilayer shown in (C), with the pattern of MoS$_2$ and WSe$_2$ indexed in green and blue colors, respectively.

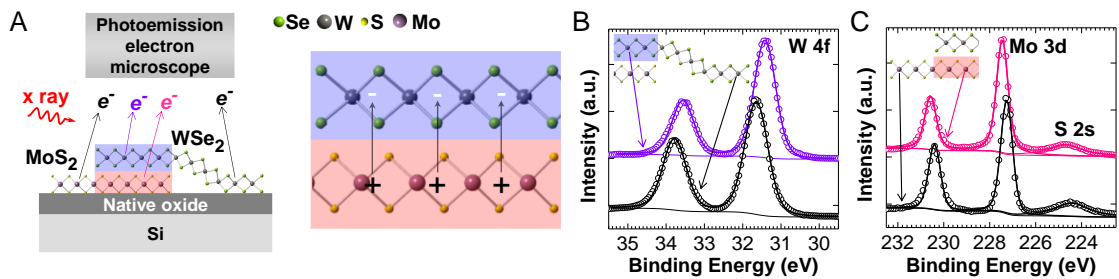

**Figure 2.** XPS core level shift analyses of WSe$_2$/MoS$_2$ heterostructures. (A) Sketch of the spatially-resolved photoelectron emission microscopy experiment. (B) Comparison of W 4f core level doublet from WSe$_2$ and WSe$_2$/MoS$_2$ indicating a 220meV shift to lower binding energy, corresponding to a negative net charge on the WSe$_2$ top layer. (C) Comparison of Mo 3d core level doublet and S 2s singlet from MoS$_2$ and WSe$_2$/MoS$_2$ indicating a shift of 190 meV to higher binding energy, corresponding to a positive net charge on MoS$_2$. The single peak at 224.4eV-224.6eV is identified as S 2s, which shows the same shift as Mo 3d, as expected.

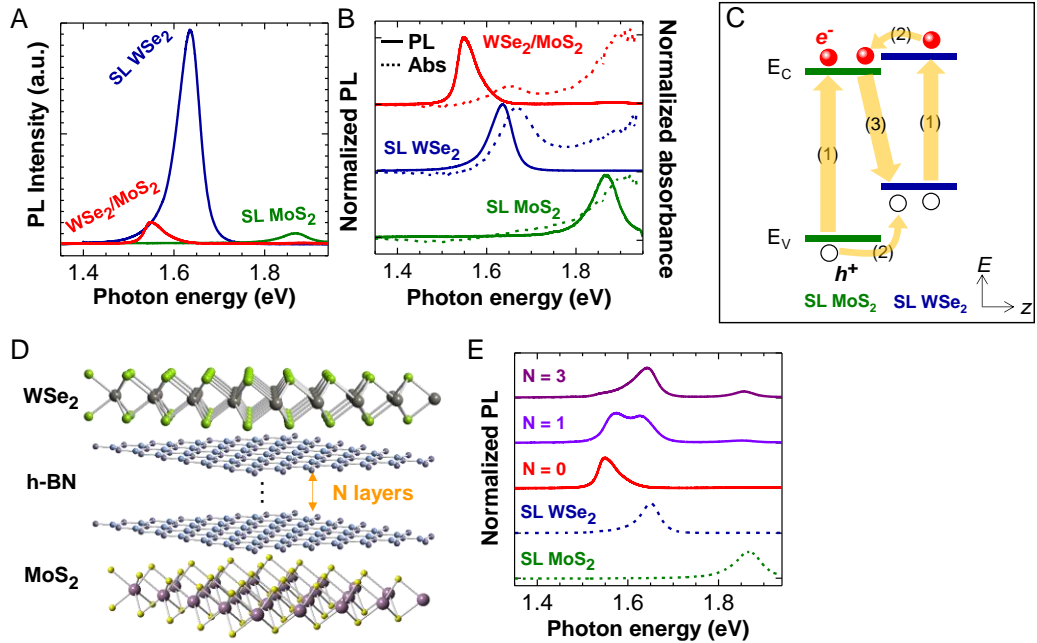

**Figure 3.** Photoluminescence and absorption from $WSe_2/MoS_2$ hetero-bilayers. (A) PL spectra of single-layer $WSe_2$, $MoS_2$, and the corresponding hetero-bilayer. (B) Normalized PL (solid lines) and absorbance (dashed lines) spectra of single-layer $WSe_2$, $MoS_2$, and the corresponding hetero-bilayer, where the spectra are normalized to the height of the strongest PL/absorbance peak. (C) band diagram of $WSe_2/MoS_2$ hetero-bilayer under photo excitation, depicting (1) absorption and exciton generation in $WSe_2$ and $MoS_2$ single layers, (2) Relaxation of excitons at the $MoS_2/WSe_2$ interface driven by the band-offset, (3) Radiative recombination of spatially indirect excitons. (D) An atomistic illustration of the heterostructure of single-layer $WSe_2$/single-layer $MoS_2$ with few-layer h-BN spacer in the vdW gap. (E) Normalized PL spectra from single-layer $WSe_2$/single-layer $MoS_2$ heterostructure with $N$ layers of h-BN ($N$=0, 1 and 3).

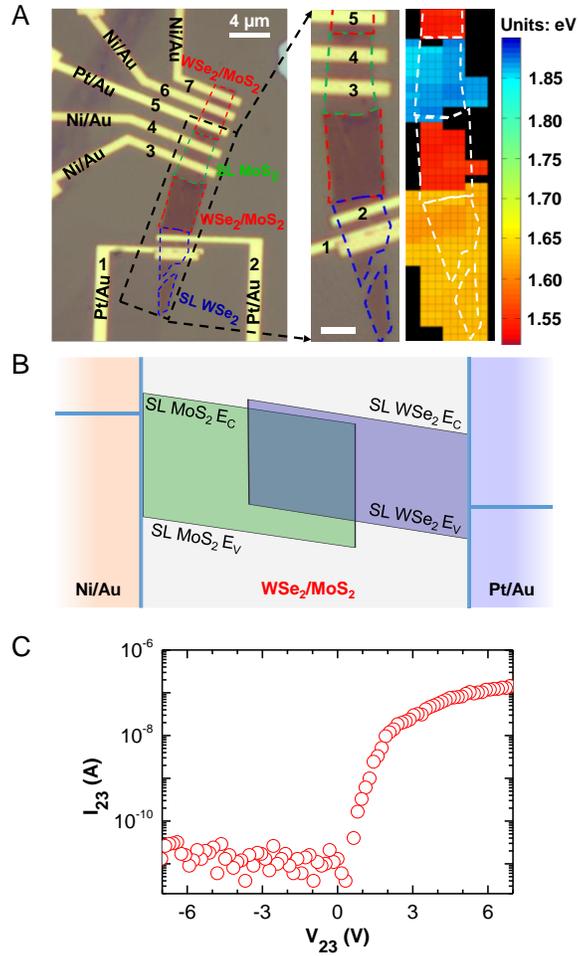

**Figure 4.** Electrical transport across the WSe$_2$/MoS$_2$ hetero-interface. (A) Optical microscope image of a device encompassing single-layer WSe$_2$, WSe$_2$/MoS$_2$ hetero-bilayer, and single-layer MoS$_2$ on a Si/SiO$_2$ substrate. Electrodes are numbered 1-7 from bottom to top. At right a color coded PL peak energy map. The scale bar is 2μm. (B) a qualitative band diagram of the single-layer WSe$_2$/hetero-bilayer/single-layer MoS$_2$ device, corresponding to the device between electrodes 2 and 3. (C) I-V characteristic when measuring between electrodes 2 and 3, with 2 grounded, 3 biased. A back gate voltage of 50 V was applied to reduce the contact resistance to MoS$_2$ while patterned NO$_2$ doping was used near the WSe$_2$ contact for reducing the contact resistance.

# Supporting Information

**Methods**

*Sample preparation:*

Fabrication of the heterostructure started with the transfer of MoS$_2$, WSe$_2$ and h-BN single and few layers on separate Si/SiO$_2$ substrates (oxide thickness, $t_{ox}$=260 nm) using adhesive tapes by the mechanical exfoliation method (1). For h-BN, a 550 nm band pass filter (FWHM=40 nm) was used to further enhance the optical contrast for locating single- to few-layer flakes on Si/SiO$_2$. These mechanically exfoliated flakes were then annealed at 250 °C for 3 hr in a H$_2$ environment (3.3 torr, 200 sccm) to remove any surface organic residues. The heterostructures were realized by a dry transfer technique (2, 3) with a PMMA membrane as the transfer media. For the heterostructures with single- or few-layer h-BN, single-layer WSe$_2$ was first transferred onto h-BN, and then these two layers together were picked up by PMMA and transferred onto single-layer MoS$_2$.

Samples for TEM were prepared on Au TEM grids with holey carbon nets of 1.2 μm diameter orthogonal hole arrays (Ted Pella, Inc.). To remove the PMMA after transfer, the grids with the heterostructure/PMMA were immersed into dichloromethane (DCM) for 6 sec, taken out to allow to air dry for 3 min in fume hoods, and then annealed in H$_2$ with the same condition noted previously. The short time of immersion is to prevent the samples escaping from the carbon net, while no blow dry was involved to keep the mechanical integrity of the relatively fragile carbon net. Samples for PEEM were prepared on natively oxidized p+ Si substrate. A similar cleaning process was used to remove PMMA and its residue as described above.

*TEM condition:*

The HRTEM image displayed in Figure 1c was acquired on a FEI Titan microscope equipped with a field emission gun operating at accelerating voltage of 300 kV. The images were recorded on a 2048x2048 Gatan Ultrascan CCD camera using DigitalMicrograph software. The microscope was operated in low-dose mode to minimize electron beam-induced degradation as well as contamination of the sample. A Wiener-type filter was applied to the images to reduce the signal arising from the amorphous background. The electron diffraction pattern shown in Figure 1d was acquired on a Zeiss Libra 200MC equipped with a monochromator and omega-type in-column energy filter using a 2048x2048 Gatan Ultrascan CCD camera. The microscope was operated at 200 kV accelerating voltage in parallel illumination mode with a semi-convergence angle of 40 μrad. The illuminated area was limited to 0.13 μm$^2$ using a condenser aperture with a diameter of 37 μm. Energy-filtered diffraction patterns were acquired and analyzed quantitatively using the dedicated software DigitalMicrograph.

*PEEM experiments:*

The photoemission electron microscopy (PEEM) experiment was conducted at the soft x-ray undulator beamline UE49-PGM-a of the BESSY-II storage ring in Berlin, Germany using elliptically polarized light at photon energies of 700 eV and 150 eV. The endstation was equipped with an Elmitec PEEM-II energy microscope/analyzer allowing energy and spatially resolved imaging. Depending on the measured sample area, fields of view between 10 and 70 µm were used resulting in a spatial resolution better than 30 and 120 nm, respectively. The total energy resolution of our measurements was 100 meV.

*Optical measurements:*

PL and Raman measurements were performed at room temperature under ambient conditions. The samples were excited with a continuous wave (cw), blue laser (473 nm) with a spot size of ~ 1 µm, unless mentioned otherwise. The original laser power was 5 mW and neutral density filters were used with optical densities of 3 and 4 (corresponding to 5 and 0.5 µW laser power on the sample). Near-IR/visible absorption measurements were performed at Beamline 1.4.3 at the Advanced Light Source using a Nicolet Magna 760 FTIR interferometer and a Nicolet Nic-Plan infrared microscope. In order to access the visible region, all measurements used a quartz beamsplitter and a silicon detector (Thorlabs, Inc.). The use of synchrotron radiation enabled a spot size of <2 µm.

*Details of the hetero-bilayer device fabrication:*

Following the fabrication of the $WSe_2/MoS_2$ hetero-bilayer, a single flake was dry etched into a long ribbon with different regions corresponding to single layer $WSe_2$, $MoS_2$, and hetero-bilayer. Pt/Au and Ni/Au metal contacts were then placed on single layer $WSe_2$ and single layer $MoS_2$ regions respectively. A $ZrO_2$ layer was deposited onto single layer $MoS_2$ region and partial hetero-bilayer region in the ribbon. The device was finally measured with $NO_2$ doping on the exposed $WSe_2$ region to minimize contact resistance at the $Pt/WSe_2$ contact, as shown in Fig. S7. Note that without $NO_2$ doping, the device still exhibits rectifying behavior with ~10x lower forward bias current density arising from the extra parasitic resistances (Fig. S8). The step-by-step fabrication process is as followed.

1. E-beam lithography was used to define etched regions in the hetero-bilayer flake. Briefly, samples were coated with PMMA (4% in chlorobenzene), baked at 180 °C for 5 min, and then exposed using an electron-beam lithography system.

2. The hetero-bilayer flakes were then patterned dry etched using $XeF_2/N_2$ gas ($XeF_2$ 3 torr, $N_2$ 1.5 torr). For 8 cycles of 30 seconds etching, the etch rate of chalcogenides is over 10 nm/min.

3. The PMMA mask was removed by a dichloromethane (DCM) wash for 10 min and the samples were then annealed in $H_2$ (3.3 torr, 200 sccm) for 3 hours to remove the PMMA residue.

4. Ni/Au (10 nm/30 nm) contacts were deposited on single layer $MoS_2$ region by an e-beam lithography, metal deposition and lift-off process. Ni/Au was chosen to contact $MoS_2$ as it is known to enable efficient injection of electrons (4, 5). A short Ni/Au "anchor" bar (~5 µm length) was also deposited at the end of $WSe_2$ side to prevent the flake detaching from the substrate during the lift-off process.

5. Pt/Au (10 nm/30 nm) contacts were deposited on single layer $WSe_2$ region by a similar e-beam lithography, metal deposition and lift-off process. Pt/Au was found to give better p-type conduction in single layer $WSe_2$ than Pd, which was the best contact to the valence band of $WSe_2$ (ref. 6).

6. A $ZrO_2$ layer (20 nm) was then deposited onto single layer $MoS_2$ region and partial hetero-bilayer region in the ribbon as a $NO_2$ blocker by an e-beam lithography, atomic layer deposition and lift-off process.

7. The fully fabricated device was then exposed to 0.05% $NO_2$ in $N_2$ gas for 10 min and measured, as shown in Fig. S7. For the I-V characteristic shown in Fig. 4C in the main text, a back gate voltage of 50 V was applied to lower the parasitic resistance from $MoS_2$ while the parasitic $WSe_2$ was still degenerately p-doped (6).

**TEM analysis**

The lattice constants of the real lattices can be calculated from

$$\frac{1}{d^2} = \frac{4}{3}\frac{h^2 + k^2 + hk}{a^2},$$

where $d$ is the length of the reciprocal lattice vector of the crystal lattice planes of (*hk0*) and $a$ is the lattice constant of the chalcogenide crystal. This calculation yields a value of (4.4±0.1)% for the lattice mismatch between the $WSe_2$ single-layer and the underneath $MoS_2$ single-layer, which is nearly identical to their bulk value. The alignment of the two lattices can also be examined with a fast Fourier transform of the two zoomed-in TEM images in Fig. 1C in the main text, as shown in Fig. S1.

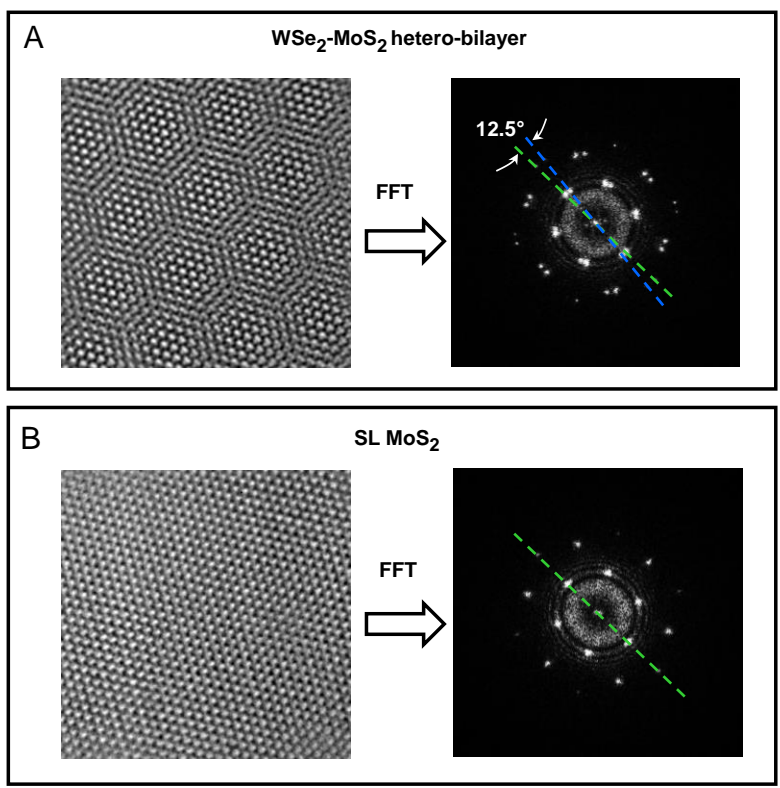

**Figure S1.** Fast Fourier transform of the HRTEM images shown in Fig. 1C in the main text for (A) WSe$_2$/MoS$_2$ hetero-bilayer. (B) single-layer MoS$_2$.

**Raman characterization of the WSe$_2$/MoS$_2$ hetero-bilayer**

There are four Raman-active modes for both WSe$_2$ and MoS$_2$, of which only A$_{1g}$, and E$^1_{2g}$ modes were observed in our measurements due to the selection rule in the back-scattering configuration and the restricted rejection against Rayleigh scattering (7). As shown in Fig. S2, the in-plane E$^1_{2g}$ mode peaks of WSe$_2$ and MoS$_2$ remained unchanged (within 0.5 cm$^{-1}$) comparing before and after transferring. This is a clear indication that there is no/negligible strain in either layer in the final hetero-bilayer, consistent with our TEM analysis. The out-of-plane A$_{1g}$ mode peak red shifted by ~ 1 cm$^{-1}$ (from 405.1 cm$^{-1}$ to 404.1 cm$^{-1}$) for MoS$_2$, while for WSe$_2$ it's hard to detect as the A$_{1g}$ peak is overlapping with the E$^1_{2g}$ peak. This small red shift should be attributed to the interlayer coupling, which is also seen in bulk/few-layer WSe$_2$ and MoS$_2$ (ref. 7, 8).

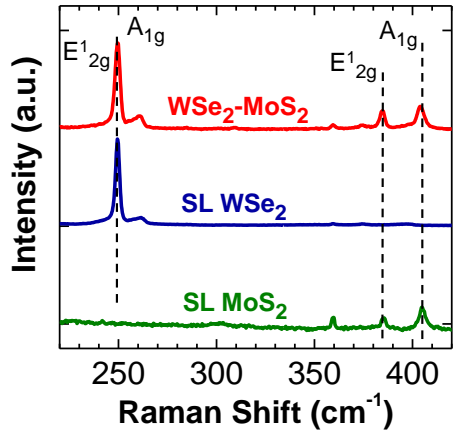

**Figure S2.** Raman spectra comparison for single-layer MoS$_2$ and WSe$_2$ and the WSe$_2$/MoS$_2$ hetero-bilayer. The excitation laser wavelength here is 532 nm.

**PEEM experiment detail**

As shown in Fig. S3A, a WSe$_2$/MoS$_2$ hetero-stack, which contained a single-layer WSe$_2$ on top of MoS$_2$ with thicknesses ranging from 1 single-layer (SL), 2 SLs and 4 SLs, was measured and analyzed by PEEM. Note that the thicknesses of the layers were determined by optical contrast, Atomic Force Microscope (AFM) in combination with PL. Figure S3B shows the W 4f$_{7/2}$ binding energy position contour, where distinct boundaries between regions can be visualized and are consistent with our sample geometry. The binding energy map was obtained by a batch fitting of all spectra collected at different detector location, single peaks of W 4f$_{7/2}$ and Mo 3d$_{5/2}$ were simulated by a Gaussian profile. For more precise analysis, signal from homogeneous parts of the image were spatially integrated and more sophisticated fitting procedure (a combination of Shirley background and Doniach-Sunjic lineshape) was used (see Figure 2). For the single-layer WSe$_2$ in contact with the substrate, the W 4f$_{7/2}$ binding energy is 33.78 eV, while for the single-layer WSe$_2$ on top of single-layer MoS$_2$, the W 4f$_{7/2}$ binding energy is 33.56 eV. The W 4f$_{7/2}$ core level binding energy decreased by ~ 220 meV when contacting to MoS$_2$, as noted in the text. The direction of the peak shift is consistent with a negative net charge on the WSe$_2$ in the WSe$_2$/MoS$_2$ hetero-bilayer as predicted by density functional theory (9). On the other hand, charge neutrality in the hetero-bilayer requires that a shift in the opposite direction is present on the MoS$_2$ component of WSe$_2$/MoS$_2$. As is shown in Fig. S3C, a shift of +190 meV with respect to stand-alone MoS$_2$ is observed indeed in the Mo 3d core levels of the WSe$_2$/MoS$_2$ (from 227.26 eV to 227.45 eV). The shifts in the W and Mo core levels are evidences that there exists charge transfer induced electric field between WSe$_2$ and MoS$_2$ layers. The hetero-bilayer can essentially be interpreted as being a two-dimensional dipole, an atomically thin parallel plate capacitor with van der Waals gap. One can also notice that the W $4f_{7/2}$ binding energy position shifted less when contacting to thicker $MoS_2$ layers by ~ 30 meV/layer. It is not clear at this stage which one of the three parameters, namely the charge transfer amount, layer distance or the dielectric constant in the van der Waals gap is playing a more important role in this slight decrease. Detailed density functional theory calculations are needed to further shed light on this issue.

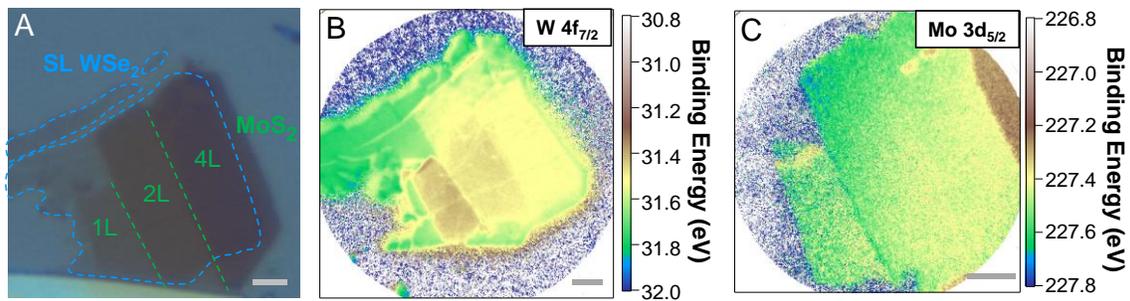

**Figure S3.** (A) The optical microscope image of the $WSe_2/MoS_2$ heterostructure for PEEM characterization. The PEEM sample was on a naturally (native) oxidized heavily p-doped Si substrate, while the image was taken when the sample was on $Si/SiO_2$ (260 nm) substrate. (B) W $4f_{7/2}$ binding energy position contour plot of the sample in (A). (C) Mo $3d_{5/2}$ binding energy position contour plot of the sample in (A). The scale bars are 2 μm.

**Statistics of PL from different WSe$_2$-MoS$_2$ hetero-bilayer samples**

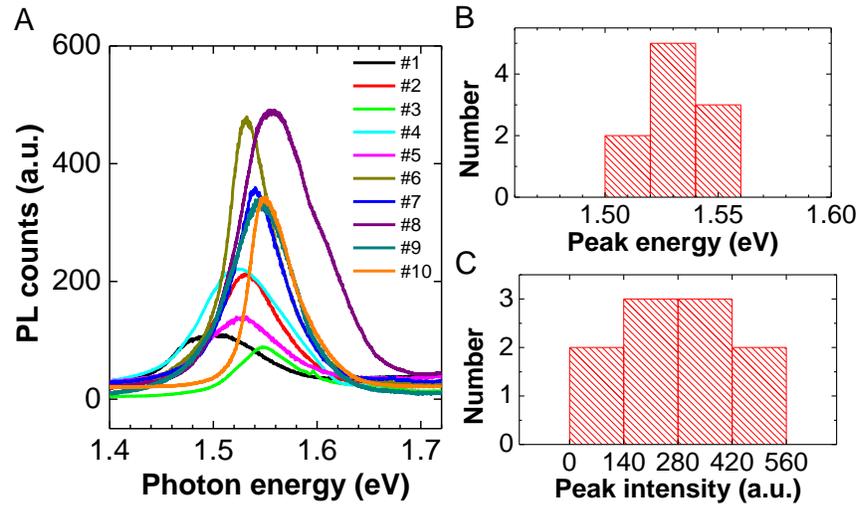

**Figure S4.** PL statistics of multiple WSe$_2$/MoS$_2$ hetero-bilayer samples. (A) PL spectra of 10 WSe$_2$/MoS$_2$ hetero-bilayers. (B) Histogram of the PL peak energies for these 10 samples. (C) Histogram of the PL peak intensities for these 10 samples.

**Luminescence tail analysis**

The band edge tail $D(v)$ as a function of frequency $v$ (also called the Urbach tail (10)), is related to the photon emission rate per unit energy by the van Roosbroeck-Shockley equation (11, 12). $D(v)$ can further be related to the photoluminescence spectrum $I(v)$ by,

$$D(v) \propto \frac{I(v)(e^{hv/kT}-1)}{n_r^2 v^2}$$

where $h$ is Planck's constant, $k$ is Boltzmann's constant, $T$ is temperature and $n_r$ is the real part of the refractive index. Figure S5 shows the shapes of the Urbach tails of single-layer WSe$_2$, single-layer MoS$_2$ as well as the WSe$_2$/MoS$_2$ hetero-bilayer. An optical semilog inverse slope of 30 meV/dec is extracted from all three tails, corresponding to the sharpness of the band edge. The nearly identical inverse slope of the hetero-bilayer proves that the band edge sharpness of the heterostructure can be as high as that of its constituent single layers.

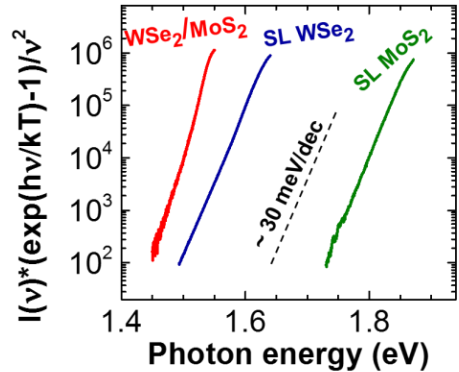

**Figure S5.** Band edge tails derived from the PL using the van Roosbroeck-Shockley equation, depicting the sharpness of the band edges. The spectra were normalized to peak magnitude.

**Absolute PL spectra of WSe$_2$/MoS$_2$ heterostructure with h-BN**

Figure S6 shows the absolute PL of WSe$_2$/MoS$_2$ hetero-stacks with single- and tri-layer h-BN spacers, along with the PL of single layer WSe$_2$. Interlayer coupling becomes negligible for the sample with a tri-layer h-BN spacer, as indicated by both the position and the intensity of the peak at 1.64 eV, which are nearly the same as single-layer WSe$_2$. The slight difference (0.01 eV) in the peak position is likely due to the fact that the boundary on the bottom side of WSe$_2$ has changed from SiO$_2$ to h-BN. On the other band, a single layer of h-BN does not fully suppress the interlayer interaction between WSe$_2$ and MoS$_2$. The PL spectrum of WSe$_2$/h-BN/MoS$_2$ shows a double peak feature centered at 1.6 eV with an intensity on the order of 1/3 of that of typical single-layer WSe$_2$. We interpret the lower energy peak component of the doublet as being the peak of the hetero-bilayer with reduced interlayer coupling due to the intercalation of the h-BN single layer. Indeed this peak falls in between the peak of the unperturbed single layer WSe$_2$ and strongly coupled WSe$_2$/MoS$_2$ bilayer. This demonstrates that the interlayer coupling can be readily tuned by intercalation of layered dielectric media into the van der Waals gap. The higher energy peak closely coincides with the peak of single layer WSe$_2$ and could be due to a competing spatially direct emission in single layer WSe$_2$ as the photo-generated electrons now see a BN barrier and would have certain probability of staying in the excitonic conduction band edge of WSe$_2$. This interpretation is also supported by the weak emission close to the position of single layer MoS$_2$.

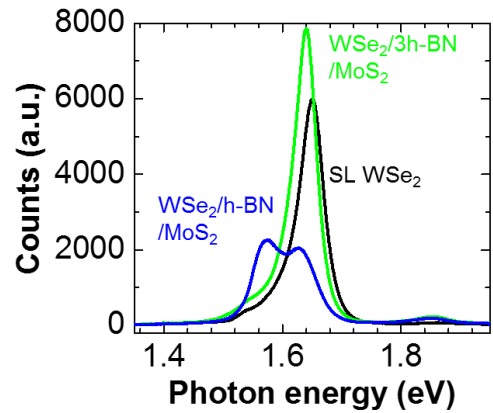

**Figure S6.** Absolute PL comparison of single-layer WSe$_2$, single-layer WSe$_2$/single-layer MoS$_2$ heterostructure with single- and tri- layer h-BN in between.

## $NO_2$ doping of $WSe_2/MoS_2$ hetero-bilayer device

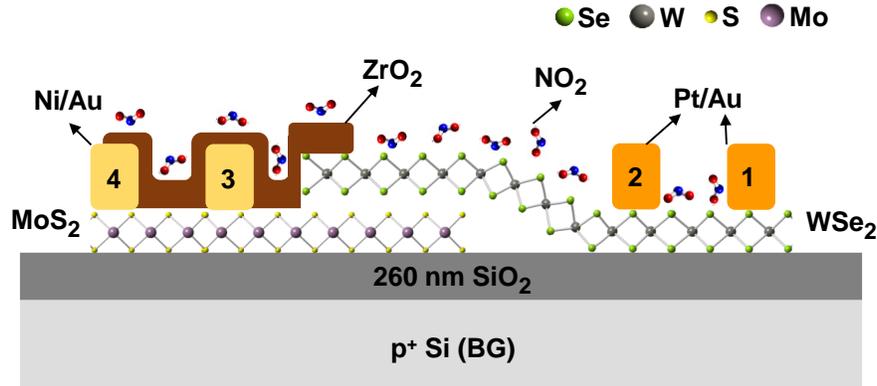

**Figure S7.** Schematic of fully fabricated $WSe_2/MoS_2$ hetero-bilayer device. $NO_2$ doping of exposed $WSe_2$ region was utilized to reduce the contact/parasitic resistances of $WSe_2$.

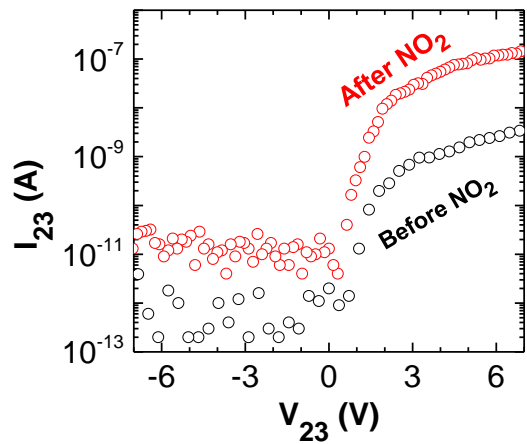

**Figure S8.** I-V characteristic when measuring between electrodes 2 and 3 (with 2 grounded, 3 biased) before and after $NO_2$ doping of the exposed $WSe_2$ region in the $WSe_2/MoS_2$ hetero-bilayer device. A back gate voltage of 50 V was applied.

**Electrical behavior of WSe$_2$, MoS$_2$ single layers**

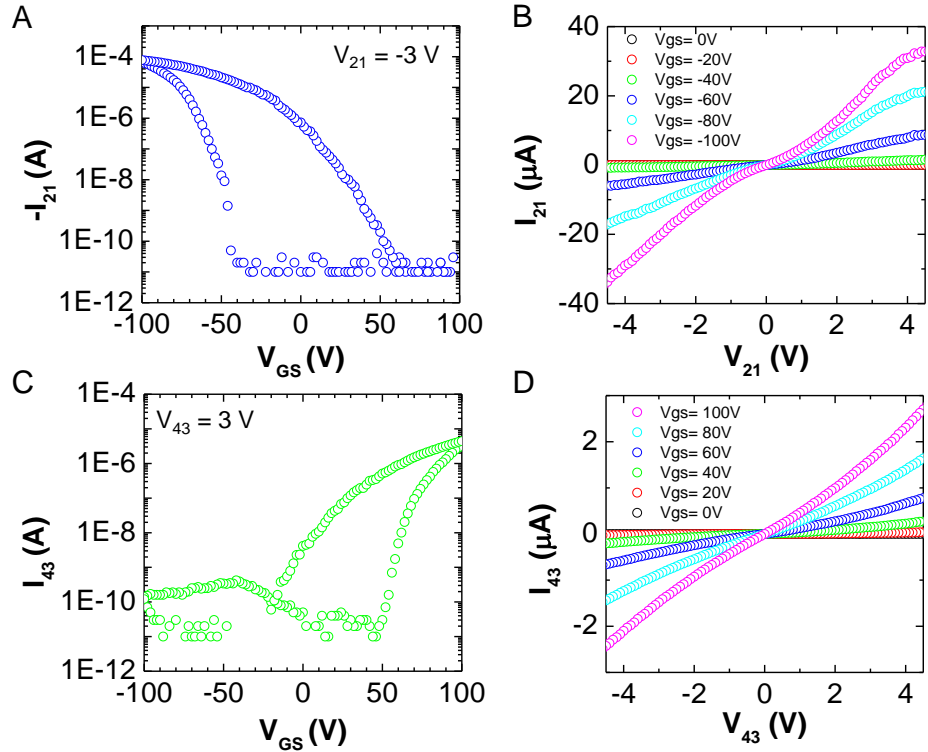

**Figure S9.** (A) Transfer characteristic of the single layer WSe$_2$ device in Fig. 4A of the main text (measured between terminal 1 and 2). (B) Output characteristic of the device in (A). (C) Transfer characteristic of the single layer MoS$_2$ device in Fig. 4A of the main text (measured between terminal 3 and 4). (D) Output characteristic of the device in (C).

**Gate dependence in the I-V characteristic of WSe$_2$/MoS$_2$ heterojunction**

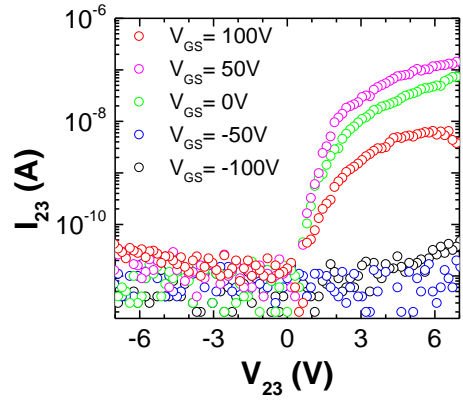

**Figure S10**. Gate dependence in the I-V characteristic of WSe$_2$/MoS$_2$ heterojunction. Due to the parasitic resistances in the WSe$_2$ and MoS$_2$ single layers, the I-V shows an "anti-ambiploar" behavior, namely the peak conductance at the forward bias region appears near an intermediate gate bias, which is similar to what has been previously observed in the MoS$_2$/nanotubes heterostructure (13).

# References

1. Novoselov KS, *et al.* (2004) Electric field effect in atomically thin carbon films. *Science* 306(5696):666-669.

2. Dean C, *et al.* (2010) Boron nitride substrates for high-quality graphene electronics. *Nat Nanotechnol* 5(10):722-726.

3. Ponomarenko L, *et al.* (2011) Tunable metal-insulator transition in double-layer graphene heterostructures. *Nat Phys* 7(12):958-961.

4. Fang H, *et al.* (2013) Degenerate n-Doping of Few-Layer Transition Metal Dichalcogenides by Potassium. *Nano Lett* 13(5):1991-1995.

5. Liu H, Neal AT, & Ye PD (2012) Channel length scaling of $MoS_2$ MOSFETs. *ACS nano* 6(10):8563-8569.

6. Fang H, *et al.* (2012) High-Performance Single Layered $WSe_2$ p-FETs with Chemically Doped Contacts. *Nano Lett* 12(7):3788-3792.

7. Zeng H, *et al.* (2013) Optical signature of symmetry variations and spin-valley coupling in atomically thin tungsten dichalcogenides. *Sci Rep* 3:1608.

8. Lee C, *et al.* (2010) Anomalous lattice vibrations of single-and few-layer $MoS_2$. *ACS nano* 4(5):2695-2700.

9. Terrones H, López-Urías F, & Terrones M (2013) Novel hetero-layered materials with tunable direct band gaps by sandwiching different metal disulfides and diselenides. *Sci Rep* 3:1549.

10. Urbach F (1953) The long-wavelength edge of photographic sensitivity and of the electronic absorption of solids. *Phys Rev* 92:1324-1324.

11. Van Roosbroeck W & Shockley W (1954) Photon-radiative recombination of electrons and holes in germanium. *Phys Rev* 94(6):1558.

12. Kost A, Lee H, Zou Y, Dapkus P, & Garmire E (1989) Band-edge absorption coefficients from photoluminescence in semiconductor multiple quantum wells. *Appl Phys Lett* 54(14):1356-1358.

13. Jariwala D*, et al.* (2013) Gate-tunable carbon nanotube–MoS$_2$ heterojunction p-n diode. *Proceedings of the National Academy of Sciences* 110(45):18076-18080.